\documentclass[aps,pra,twocolumn,superscriptaddress,floatfix]{revtex4-2}
\usepackage[dvipdfmx]{graphicx,xcolor}
\usepackage{bm} 
\usepackage{color} 
\usepackage{amsmath}
\begin{document}
\title{
Cold-atom fountain for atom-surface interaction measurements mediated by a near-resonant evanescent light field}
\author{Taro Mashimo}
\affiliation{Department of Physics, Faculty of Science and Engineering,
Chuo University, Kasuga, Bunkyo-ku, Tokyo 112-8551, Japan}
\author{Masashi Abe}
\affiliation{Department of Physics, Faculty of Science and Engineering,
Chuo University, Kasuga, Bunkyo-ku, Tokyo 112-8551, Japan}
\author{Athanasios Laliotis}
\affiliation{Laboratoire de Physique des Lasers, Universit\'e Sorbonne Paris Nord, F-93430 Villetaneuse, France}
\affiliation{CNRS, UMR 7538, LPL, 99 Avenue J.-B. Cl\'ement, F-93430 Villetaneuse, France}
\author{Satoshi Tojo}
\affiliation{Department of Physics, Faculty of Science and Engineering,
Chuo University, Kasuga, Bunkyo-ku, Tokyo 112-8551, Japan}
%
%
\date{\today}

\begin{abstract}
Cold atomic ensembles offer precise tools for probing near-field interactions, yet experimental data linking atom dynamics to surface-induced forces remains limited. 
This study investigated the interaction between atoms and a dielectric surface using an atomic fountain measurement technique, in which cold rubidium atoms were released from a moving optical dipole trap. 
The launched cold atoms were irradiated with an evanescent light detuned from the D$_2$ transition by $-$20.2 to $+$20.2 MHz, after which they were recaptured by reactivating the optical dipole trap. 
Our measurements revealed that the number of recaptured atoms decreased with increasing flight time, and the decay was suppressed under blue-detuned conditions. 
We modeled the motion dynamics of the cold atomic ensemble, incorporating Casimir-Polder interactions between the dielectric surface and cold atoms, and observed that the rate of decrease in the number of residual atoms depended on the value of the van der Waals potential coefficient $C_3$. The calculation results demonstrated good agreement with the experimental results, allowing us to estimate $C_3 = 5.6^{+2.4}_{-1.9} \times 10^{-49}$ Jm$^3$ by comparing simulations with the experimental results across various $C_3$ values, accounting for experimental errors.
\end{abstract}

%
%
%
%

\maketitle

\section{Introduction}
In recent years, quantum sensors based on dilute atomic ensembles have found widespread application in probing both classical and quantum mechanical phenomena \cite{degenQuantumSensing2017b}. 
While numerous quantum sensors employ solids, liquids, or quasiparticles within solids, these platforms often face challenges related to robustness and operability owing to their strong coupling with external fields \cite{ohtsuNearFieldNanoOptics1999}. 
In contrast, gas probes employing dilute neutral atomic gases present significant advantages for investigating unknown fields. 
Notably, the study of low-density gases facilitates the elucidation of novel physical phenomena, as many systems can be reduced to two-body problems, such as cold atoms and Bose-Einstein condensates \cite{weinerExperimentsTheoryCold1999}.

Atomic fountains facilitate the prolonged interaction of atoms within an objective target region by exploiting the gravitational potential, which acts in the downward direction \cite{claironRamseyResonanceZacharias1991, croninOpticsInterferometryAtoms2009a}. 
This extended interaction time allows for enhanced atom-atom interactions and underpins a broad range of applications, including high-sensitivity measurements approaching the standard quantum limit through quantum interference, as well as measurements of fundamental physical constants through prolonged coherence times \cite{croninOpticsInterferometryAtoms2009a}. 
By employing launched atomic systems, the initial velocity and position of the atoms can be precisely controlled, thereby maintaining their proximity to surfaces for an extended duration. Cold atomic gases, distinguished by their substantially reduced momentum, offer superior spatial manipulability and resolution compared with room-temperature gases. 
The use of atoms as sensors for probing the near-field properties of dielectric materials is in an increasingly prominent area of research \cite{laliotisAtomsurfacePhysicsReview2021}. 
Extensive experimental studies on the fall of cold atoms or Bose-Einstein condensates interacting with evanescent light fields at dielectric surfaces are underway: 
These studies encompass dielectric surfaces \cite{kasevichNormalincidenceReflectionSlow1990, bongsCoherentEvolutionBouncing1999}, measurement of the van der Waals force \cite{landraginMeasurementVanWaals1996}, observation of the Casimir-Polder effect \cite{benderDirectMeasurementIntermediateRange2010}, trapping of cold atoms using evanescent light \cite{rychtarikTwoDimensionalBoseEinsteinCondensate2004a},
diffraction of Bose-Einstein condensates on metallic gratings \cite{benderProbingAtomSurfaceInteractions2014a},
the generation of electromagnetically induced transparency in optical nanofibers \cite{kumarAutlerTownesSplittingFrequency2015}, 
and atomic fountain clocks \cite{lewisGratingchipAtomicFountain2022}.
%
%
\begin{figure}
\begin{center}
\includegraphics[width=7.0cm]{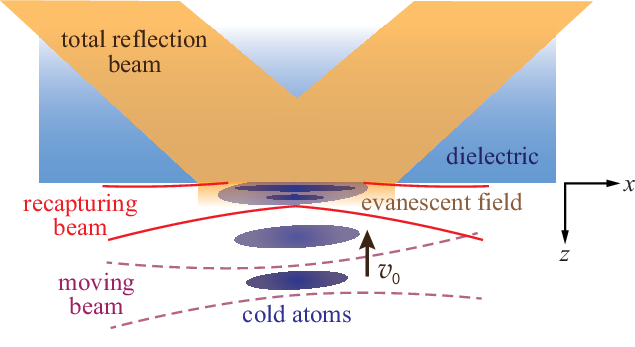}
\caption{
Geometry of the cold atomic fountain measurement. 
The cold atoms transported by the optical dipole trap as the moving beam are launched upward with an initial velocity $v_0$ by switching off the trap beam. 
When light undergoes total internal reflection at the dielectric interface, it generates an evanescent field at the dielectric interface that interacts with the launched cold atoms. 
After traversing the evanescent region, the atoms are recaptured by the optical dipole trap, now functioning as the recapturing beam. 
This recaptured beam, when reflected at the dielectric surface, forms a standing wave that facilitates efficient recapture.
}
\label{geoexp}
\end{center}
\end{figure}

In this study, optically trapped cold Rb atoms were transported near a dielectric surface, enabling surface exploration using a cold atomic fountain. 
As shown in Fig.~\ref{geoexp}, the atoms are released with an upward initial velocity $v_0$ and subsequently exposed to an evanescent light field near the resonant-frequency detuning. 
The interaction between the cold atoms and dielectric surface is probed as the atoms traverse the evanescent field, after which they are recaptured by the optical dipole trap.
The released cold atoms that collide with the dielectric 
surface in the interaction region are irretrievably lost from the trap, as they acquire significant energy from the room-temperature dielectric.
The number of recaptured atoms decreases owing to attractive interactions arising from the near-field Casimir-Polder (van der Waals) potential between the dielectric surface and ground state of Rb, which is proportional to $z^{-3}$, where $z$ represents the distance from the surface.
When atoms traverse an evanescent field with either blue- or red-detuned frequencies, they experience surface-induced interactions—repulsive for blue detuning and attractive for red detuning. 
The influence of these surface interactions is analyzed by comparing the difference in the number of recaptured atoms in both the experimental and calculated results.

\section{Theory}
\subsection{Optical dipole and radiative forces of an evanescent field}
The trap beam potential in free space is described using the Gaussian beam optics theory \cite{mandelOpticalCoherenceQuantum1995}.
Atom-light interaction generates the optical dipole and radiative forces \cite{lettOpticalMolasses1989};
\begin{eqnarray}
\bm{F}_{\rm{D}} &=& \frac{\hbar \delta}{4} \frac{\nabla s'}{1+s'+\delta^2/(\Gamma/2)^2}, \\
\bm{F}_{\rm{R}} &=& \frac{\hbar \bm{k} \Gamma}{2} \frac{s'}{1+s'+\delta^2/(\Gamma/2)^2},
\label{diprad}
\end{eqnarray}
where $\delta$ denotes the frequency detuned from the resonance, $s' = I / I_s$ denotes the saturation parameter, $\Gamma$ denotes the natural width of the resonance, and $I$ and $I_s$ denote the beam intensity and the saturation intensity, respectively. 
In free space near resonance, a comparison between the dipole and radiative forces reveals that the dipole force is typically negligible relative to the radiative force. 
This is because $|\nabla s'|$, which is proportional to the gradient of the beam waist, is considerably smaller than the wavelength modulus $|\bm{k}|$ in free space.
However, within the evanescent field, a pronounced gradient of the electric field exists at the dielectric-vacuum interface.
The $z$ and $x$-axes were defined as perpendicular and parallel to the surface, respectively, as shown in Fig.~\ref{geoexp}.
The intensity parameter of the evanescent field is expressed as $s_z' = s_0' e^{-2k_z z}$
whereas that of the propagating wave is $s_x' = s_0' e^{2ik_x x}$.
where $k_z = k_0 \sqrt{n^2 \sin^2\theta-1}$ and $k_x = k_0 n\sin\theta$ are the $z$ and $x$-components of the wave vector of the evanescent field, respectively, 
$s_0$ denotes the saturation parameter on the surface ($z=0$), $n$ denotes the refractive index of the dielectric, $k_0$ denotes the wave vector in vacuum, and $\theta$ denotes the angle of incidence of the total reflected beam.
Therefore, the dipole and radiative forces of the evanescent field can be expressed as follows:
\begin{eqnarray}
\bm{F}_{\rm{ED}} &=& \frac{\hbar \bm{k}_z \delta}{2} \frac{s_0' e^{-2k_z z}}{1+s_0' e^{-2k_z z}+\delta^2/(\Gamma/2)^2}, \label{evdip}\\
\bm{F}_{\rm{ER}} &=& \frac{\hbar \bm{k}_x \Gamma}{2} \frac{s_0' e^{-2k_z z}}{1+s_0' e^{-2k_z z} +\delta^2/(\Gamma/2)^2},
\label{evrad}
\end{eqnarray}
where $\bm{k}_x$ and $\bm{k}_z$ represent the wave vectors of the evanescent field in the $x$ and $z$-directions, respectively.
The ratio of the dipole force to the radiative force in the evanescent field can be expressed as follows:
\begin{equation}
R_{\rm{DR}} = \frac{|\bm{F}_{\rm{ED}}|}{|\bm{F}_{\rm{ER}} |} = \frac{k_z \delta}{k_x \Gamma} = \frac{\sqrt{n^2\sin^2\theta-1}}{n \sin\theta} \frac{\delta}{\Gamma}.
\label{ratedr}
\end{equation}
The ratio $R_{\rm{DR}}$ indicates that the dipole force perpendicular to the interface in the evanescent field is comparable with the radiative force in the propagation direction, even near the resonance regime.

\subsection{Optical dipole and radiative forces of the evanescent field with near-field Casimir-Polder potential}

In close proximity to a dielectric surface, the influence of the evanescent field modified by the near-field Casimir-Polder potential from the surface should be considered. 
The scattering force of the evanescent field with a near-field Casimir-Polder potential
$U(z) = -C_3/z^3$ can be expressed as \cite{groverPhotoncorrelationMeasurementsAtomiccloud2015}.
Generally, the optical dipole and radiative forces can be reformulated from Eqs. ~(\ref{evdip}) and (\ref{evrad}) as follows:
\begin{eqnarray}
F_{\rm{ED}} &=& \frac{\hbar k_z \delta(z)}{2} \frac{s_0' e^{-2k_z z}}{1+s_0' e^{-2k_z z}+[\delta(z)]^2/(\Gamma/2)^2}, \\ 
F_{\rm{ER}} &=& \frac{\hbar k_x \Gamma}{2} \frac{s_0' e^{-2k_z z}}{1+s_0' e^{-2k_z z} +[\delta(z)]^2/(\Gamma/2)^2},
\label{evdrvdW}
\end{eqnarray}
where $\delta(z) = \delta + C_3/(\hbar z^3)$ and $C_3$ denotes the van der Waals coefficient. 
The total force along the $z$-axis comprises both the dipole and near-field Casimir-Polder potentials, excluding the effect of gravity.
The ratio of the $z$-and $x$-directional forces in Eq. ~(\ref{ratedr}) assuming that $s_0' \gg 1$ can be expressed as follows:
\begin{eqnarray}
R_{Fz/Fx} &=& \frac{F_{\rm{ED}}+F_{\rm vdW}}{F_{\rm{ER}} } \simeq \frac{\hbar k_z \delta(z) - 6C_3/z^4}{\hbar k_x \Gamma} \nonumber \\
&=& \frac{\sqrt{n^2\sin^2\theta} \cdot \delta(z) - 6C_3/(\hbar k_0 z^4)}{n \sin\theta \cdot \Gamma}.
\label{ratedrc3}
\end{eqnarray}

\subsection{Calculation model}

In the experimental setup, the atomic gas was released in the $-z$-direction with initial position $z_0$ and velocity $v_0$.
The accelerations along the $z$- and $x$-axes, resulting from the interaction between the atoms and surface, are expressed as follows:
\begin{eqnarray}
\!\!\!\!\!\!\!\!\!\!a_{z}(z) &=& g + \frac{\hbar k_z \delta(z)}{2m} \nonumber \\
&&\!\!\!\!\!\!\!\!\times\frac{s_0' e^{-2k_z z}}{1+s_0' e^{-2k_z z}+[\delta(z)]^2/(\Gamma/2)^2} -\frac{3C_3}{mz^4}, \\ 
\!\!\!\!\!\!\!\!\!\!a_{x}(z) &=& \nonumber \\
&&\!\!\!\!\!\!\!\!\!\!\!\!\!\!\!\!\!\!\frac{\hbar k_x \Gamma}{2m} \frac{s_0' e^{-2k_z z}}{1+s_0' e^{-2k_z z} +[\delta(z)-k_xv_x]^2/(\Gamma/2)^2},
\label{accelzx}
\end{eqnarray}
where $m$ represents the mass of the Rb atom and $k_x v_x$ represents the Doppler shift. 
The atomic velocities and positions of atoms were subsequently calculated $in situ$ using classical mechanics. 
Atoms eligible for capture by the recapturing beam must satisfy two criteria: their positions must lie within the beam region, and their kinetic energies must not exceed the capture energy threshold.

In the context of experimental parameters, the atomic ensemble 
$n(x, z, v_x, v_z, t)$ at flight time $t$ after release is assumed to have a step-function spatial distribution of length $2l$. 
The center position $z_0$ and half-length $l$ are determined based on experimental conditions, with estimated values of $z_0 = 57$ $\mu$m and $l = 30$ $\mu$m, respectively. 
The initial velocity distribution of the atomic ensemble, corresponding to a temperature of $T = 22$ $\mu$K, follows the Maxwell-Boltzmann distribution. 
The initial velocity of each atom is given by the sum of the initial velocity $v_0 = -44$ mm/s and velocity distribution.
At the time of recapture, the following parameters were defined:
threshold $z_{\rm th}$ of the recapturing beam and 
velocity threshold $v_{z\rm th}$ in the $z$-direction, as well as the Rayleigh length $x_0$ of the recapturing beam and velocity threshold $v_{x\rm th}$ in the $x$-direction. 
The number of recaptured atoms is expressed as follows:
\begin{eqnarray}
N(t) = && \nonumber \\
&&\!\!\!\!\!\!\!\!\!\!\!\!\!\!\!\int_{-x_0}^{x0}\!\left[\int_0^{z_{\rm th}} \!\!\!\!\!\!\!n(x,z,v_x,v_z,t) u(v_{z\rm th}\!\!-\!|v_z|)dz\right]\!\!u(v_{x\rm th}\!\!-\!|v_x|)dx.
\nonumber \\
\label{ntot}
\end{eqnarray}
where $u(v)$ denotes a unit step function.
In the experiment, photon scattering resulting from stray light associated with the total reflection beam was observed, along with slight absorption by ordinary light.

\section{Experimental Apparatus}

%
%
\begin{figure}
\begin{center}
\includegraphics[width=7.0cm]{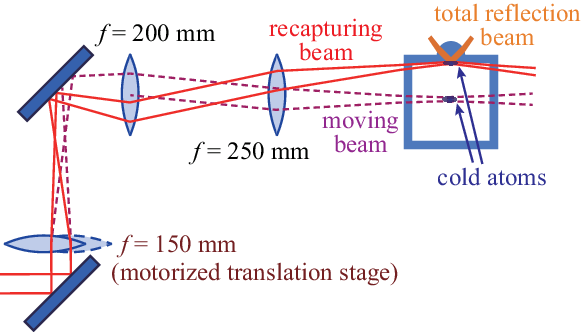}
\caption{
Experimental geometry of the velocity-controlled system for cold atoms using an optical dipole trap. 
The position of the lens, mounted on a motorized translation stage, is adjusted horizontally to control the vertical position of cold atoms confined by the moving beam. 
After being transported by the moving beam, the cold atoms are released and subsequently recaptured by the recapturing beam by turning on the identical beam following a designated flight time.
}
\label{geostage}
\end{center}
\end{figure}

The experimental geometry is shown in Fig.~\ref{geostage}.
Initially, pre-cooled Rb atoms were prepared at 10 $\mu$K using polarization gradient cooling and gray molasses technique \cite{shibataLoadingAtomsOptical2017}. 
Subsequently, the atoms were loaded into a near-resonant optical dipole trap (wavelength: 782 nm, power: 60 mW, and trap depth: 250 $\mu$K) \cite{mashimoEffectiveTrappingCold2019}.
The optical dipole trap beam is focused by a lens that is translated perpendicularly to the beam axis via the motorized translation stage.
Following the capture of atoms near the focal point of the optical dipole force trap, the focus position was vertically adjusted to control the atomic trajectory.
The atoms were transported to the dielectric surface within approximately 400 ms.
Cold atoms transported near the surface experience optical pumping owing to the presence of a trapping beam that is nearly resonant.
The transition to the ground state $F = 2$ is observed under conditions where the number of atoms is $2 \times 10^6$ and the temperature is 22 $\mu$K.
At this instant, the atoms were transported to $z_0 =$ 57 $\mu$m from the dielectric surface with a transport velocity of  $v_0 = -44$ mm/s,
and the trap beam was deactivated to release the atoms under the condision of a velocity distribution at 22 $\mu$K.
Some atoms approached the dielectric surface composed of the TEMPAX Float glass with a refractive index of 1.467 at 780 nm.
The dielectric is composed of glass and is bonded to a hemispherical lens through an optical contact.
The inclination angle between the $x$-axis direction and horizontal plane was 1.5$^{\circ}$, an effect considered negligible for the experiment.
Light incident from the hemispherical side generates evanescent light at the dielectric-vacuum chamber interface.
The incident angle is $50 \pm 1.5^{\circ}$, resulting in an evanescent field with an intensity of $I = 14.6 I_s$ where $I_s$ represents the saturation intensity of 3.577 mW/cm$^2$ for linearly polarized light resonant with the Rb D$_2$ ($F=2 - F'=3$) transition.
The frequency of the evanescent light is tunable from $-$20.2 to $+$20.2 MHz relative to resonance.
The interval between the release and recapture of the atoms, defined as flight time, ranged from 0 to 2.5 ms.
Recaptured atoms were held for another 2.5 ms to allow for thermal equilibrium. 
The recaptured atoms were then released, and after a time-of-flight of 0.02 ms, the number of atoms was measured by irradiating them with probe light for 0.1 ms using absorption imaging. 

%
%
\begin{figure}
\includegraphics[width=7.0cm]{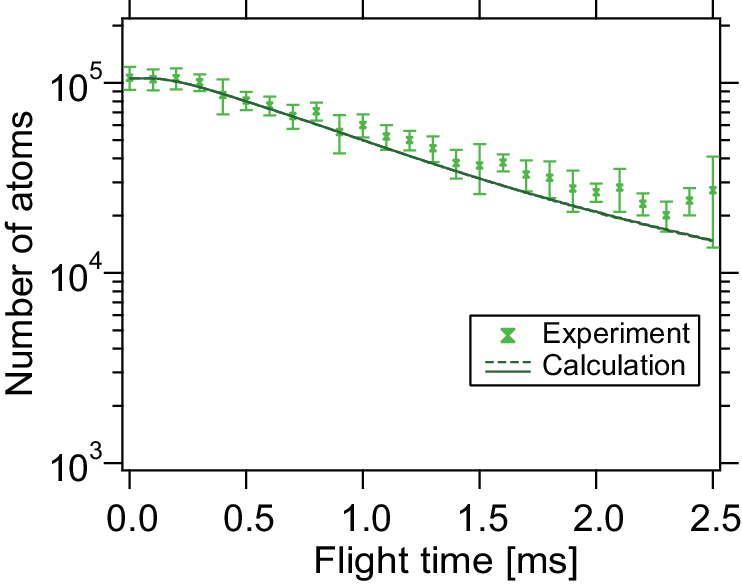}
\caption{Flight time dependence of the recaptured number of atoms. 
Experimental results (symbols) are compared with calculations using $C_3 = 5.6\times 10^{-49}$ Jm$^3$ (solid line) and $C_3 = 0$ (dashed line). }
\label{expfountain}
\end{figure}
%

%
%
\begin{figure*}
\includegraphics[width=14.0cm]{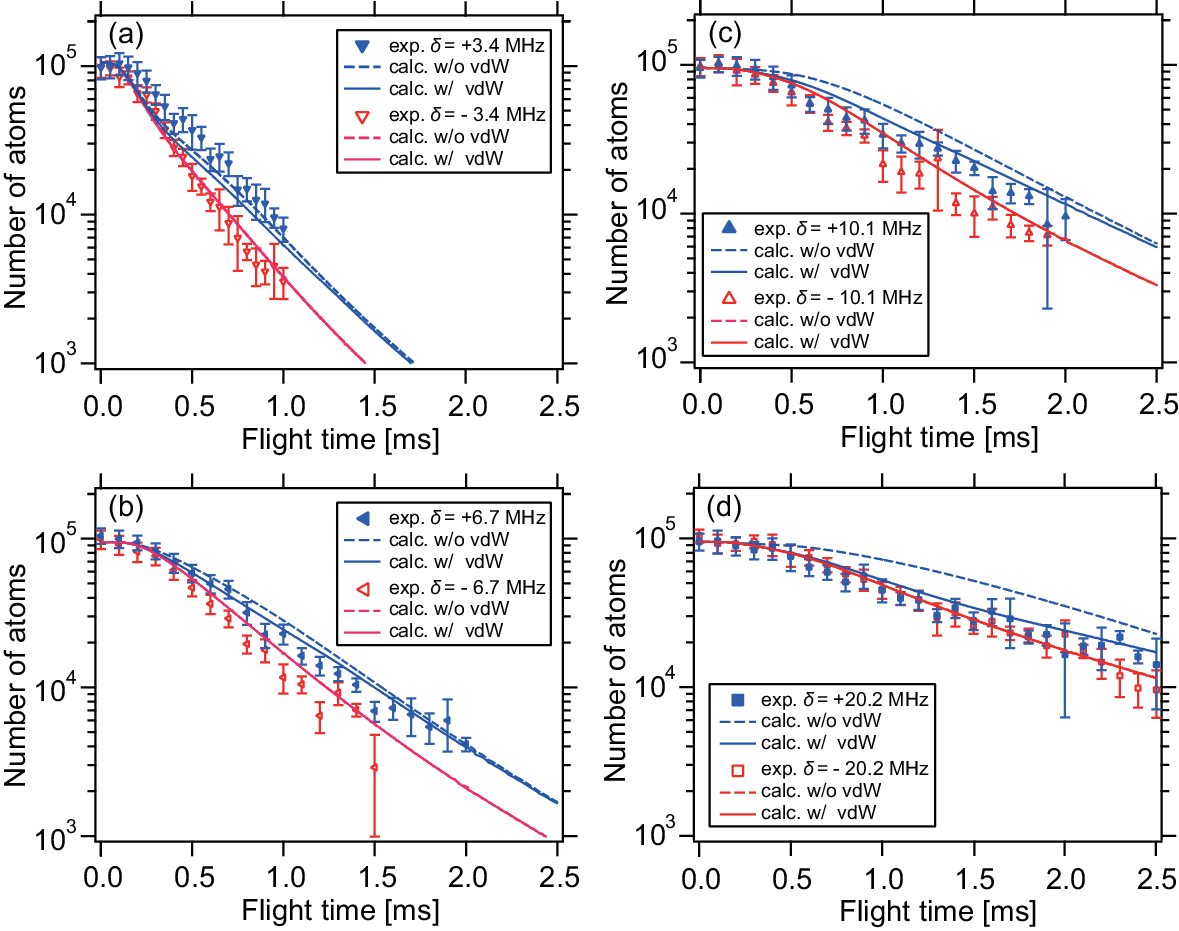}
\caption{Flight time dependence of the recaptured number of atoms. Experimental results (symbols) are compared with the calculations using $C_3 = 5.6\times 10^{-49}$ Jm$^3$ (solid line) and $C_3 = 0$ (dashed line); (a) frequency detuning $\delta = \pm 3.4$ MHz, (b) $\delta = \pm 6.7$ MHz, (c) $\delta = \pm 10.1$ MHz, and (d) $\delta = \pm 20.2$ MHz of the evanescent light.}
\label{expfountaineva}
\end{figure*}

The cold atomic fountain procedure was performed according to the following protocol:
As shown in Fig.~\ref{geoexp}, the cold atoms, confined within a vertically upward-moving optical dipole force trap beam, referred to as the moving beam, are ejected in an upward direction.
The cold atoms were guided through a transient field induced on a dielectric surface. 
Subsequently, they were recaptured by the recapturing beam, which was identical to the moving beam when activated. 
Finally, the cold atoms were detected using absorption imaging on a CCD camera.
The atomic number was determined by analyzing the variation in flight time between their release and recapture.

\section{Results and discussion}

The results of the cold atomic fountain experiment without evanescent-light irradiation are shown in Fig.~\ref{expfountain}.
Following the initiation of the process, the number of atoms remained relatively stable, with a slight decrease observed at approximately 0.3 ms.
The trap beam was incident on the dielectric surface at an angle of 3.5$^{\circ}$ to the horizontal plane (corresponding to an incident angle of 86.5$^{\circ}$) and was $p$-polarized.
Near the surface, the trap depth increased fourfold, resulting in standing waves, enabling the confinement of atoms with higher energies compared with conventional dipole traps.
Conversely, atoms external to the standing wave were also trapped within the trap beam, as the potential depth remained sufficient.
Cold atoms reaching the surface collided with the dielectric surface at a temperature matching that of the environment. 
Given that the room temperature of the glass was significantly higher than that of the atomic cloud, we assumed that all atoms that collided with the surface were lost.
After such collisions, atoms were presumed to acquire velocities exceeding the trap depth, thereby facilitating their escape from the trapping region.
By utilizing Eqs.~(\ref{accelzx}) and (\ref{ntot}) with $C_3 = 0$ value, the atomic ensemble was modeled as a cylindrical distribution centered at $z_0 = 57$ $\mu$m with a half width of $l = 30$ $\mu$m. Furthermore, the initial velocity distribution was characterized by a beam velocity of 44 mm/s and temperature of $T = 22$ $\mu$K.
The trajectories and time evolutions of the atomic momentum were calculated $in situ$.
Assuming that atoms with energies below the trap depth were recaptured within the trap beam region, the number of remaining atoms was determined using the threshold values $z_{\rm th} = 12.8$ $\mu$m for the standing wave and 100 $\mu$m for the optical dipole trap, $x_{\rm 0} = 1.5$ mm at the Rayleigh length, and threshold potential depths of 12 $T$ and 5 $T$. The potential depth of the standing wave was averaged over the recapture period. 
The calculated numbers of recaptured atoms are shown as solid lines in Fig.~\ref{expfountain}, demonstrating good agreement with the experimental results.
For comparison, the results obtained by incorporating $C_3 = 5.6 \times 10^{-49}$ Jm$^3$ in Eqs.~(\ref{accelzx}) and (\ref{ntot}) are indicated as dashed lines in Fig.~\ref{expfountain}.
The calculations with and without the $C_3$ term yield nearly identical results.
This comparison indicates that cold atomic fountain experiments aimed at probing near-field Casimir-Polder interactions in the absence of an evanescent field require more precise measurement techniques.

Subsequently, cold atomic fountain experiments were performed using an evanescent light field near resonance.
The experimental results as the frequency detuning of the evanescent light was varied, are shown in Fig.~\ref{expfountaineva}, with 
(a) $\delta = \pm 3.4$ MHz, (b) $\delta = \pm 6.7$ MHz, (c) $\delta = \pm 10.1$ MHz, and (d) $\delta = \pm 20.2$ MHz.
A substantial decrease in the number of recaptured atoms was observed when the frequency approached the resonant frequency. 
This phenomenon could be attributed to the scattering effect: the total reflected light, which generated the evanescent field, underwent multiple internal reflections within the hemispherical lens, and then irradiated the entire atomic ensemble as stray light.
Under resonant conditions, we estimated the photon scattering rate to be 0.98 ms$^{-1}$.
A comparison of experiments with equal absolute detuning from resonance, but with opposite signs, revealed that the number of trapped atoms was greater for positive detuning than that for negative detuning experiments. 
Furthermore, as the absolute value of detuning increased, the difference in the number of atoms between positive and negative detuning diminished.

The calculation results for $C_3 = 0$, assuming no effect of the near-field Casimir-Polder potential (dashed line), and for $C_3 = 5.6 \times 10^{-49}$ Jm$^3$ (solid line), including the effect of the near-field Casimir-Polder potential, are shown in Fig.~\ref{expfountaineva}.
As demonstrated in Eqs.~(\ref{ratedr}) and (\ref{ratedrc3}), even in near resonance, the radiative and dipole forces are of comparable magnitudes in the evanescent field. 
While the dipole force can typically be neglected in free space, it becomes significant in this context. 
The dipole force was predominant in the $z$-direction, perpendicular to the dielectric surface, whereas the radiative force was dominant in the horizontal $x$-direction.
We performed $in situ$ calculations assuming that only atoms with energies lower than the trap depth and confined within the trap region could be captured.

As indicated by the dashed lines in Fig.~\ref{expfountaineva}, 
we observed that negative detuning of the evanescent light led to a decrease in the number of trapped atoms compared to the case of positive detuning, where the atoms were repelled from the surface. 
For negative detuning, our theoretical predictions reproduced the experimental data very well, and the influence of the Casimir–Polder interaction was minimal. 
This was because the attractive atom–surface potential only slightly enhanced the effect of the already attractive evanescent field. 
In contrast, for positive detuning---where the Casimir–Polder interaction opposed the repulsive effect of the evanescent wave---its contribution becomed evident, particularly at large detunings as shown in Figs.~\ref{expfountaineva}(c) and (d). 
This behavior was accurately captured by our theoretical model, allowing us to perform a quantitative measurement of the Casimir–Polder potential between ground-state Rb atoms and the glass surface.
%

%
%
\begin{figure}
\includegraphics[width=7.0cm]{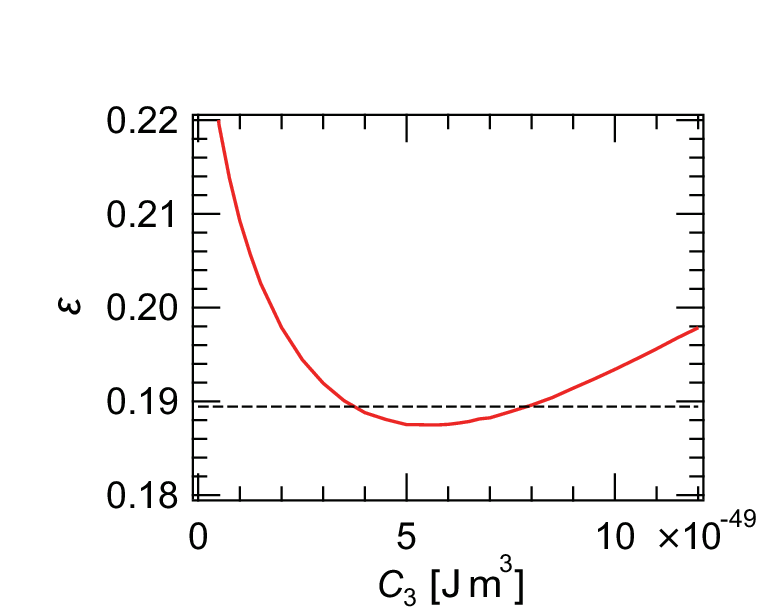}
\caption{Least square errors between experiments and calculations with $C_3$. 
The dashed line is determined using experimental and systematic errors.}
\label{c3fit}
\end{figure}

To further quantify this effect, the $C_3$ coefficient of the near-field Casimir-Polder potential was varied and the results were compared with experimental data, as shown in Fig.~\ref{c3fit}, which presents the standard deviation of the weighted experimental values.
Considering the fluctuations in the experimental number of atoms and uncertainties in the angle of total internal reflection, a systematic error of approximately 0.011 was expected.
The estimated van der Waals coefficient $C_3 = 5.6^{+2.4}_{-1.9} \times 10^{-49}$ Jm$^3$, 
which closely matches our precisely calculated value of $C_3 = 5.33 \times 10^{-49}$ Jm$^3$ for the ground state of Rb atoms. This calculation assumes a refractive index of 1.467 and utilizes the dipole moment matrix elements reported in \cite{heavensRadiativeTransitionProbabilities1961a, safronovaRelativisticManybodyCalculations1999}. 

%
%
\begin{figure}
\includegraphics[width=7.0cm]{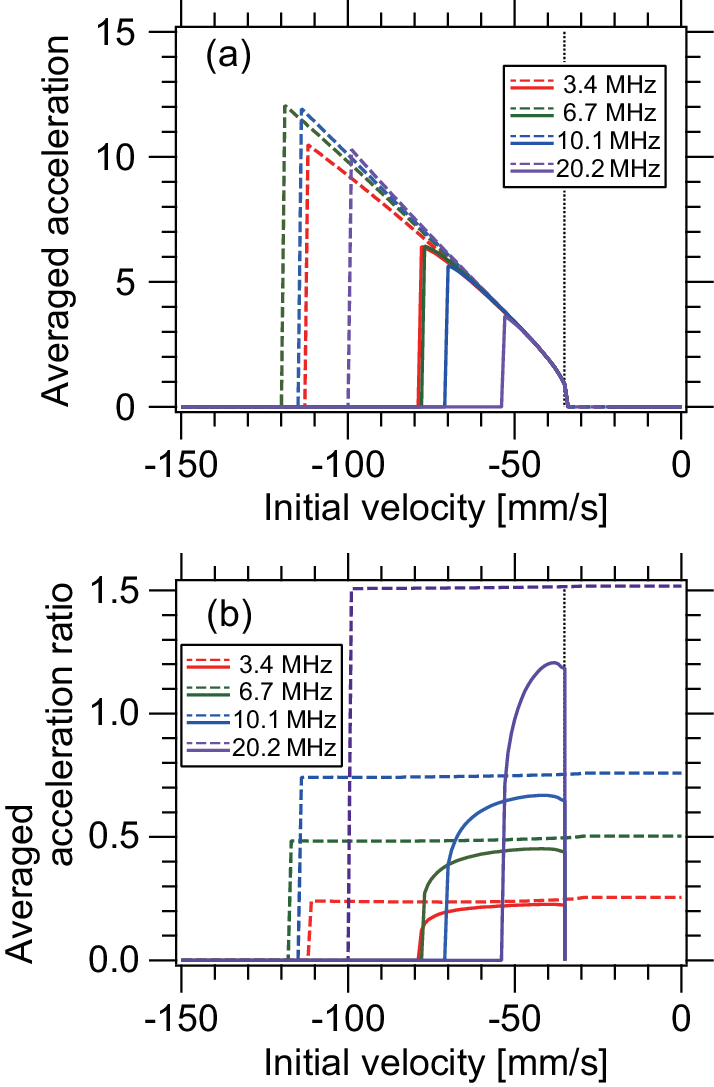}
\caption{(a) Averaged acceleration $(a_z-g)/g$ with
$C_3 = 0$ (dashed lines), and $C_3 = 5.6\times10^{-49}$ Jm$^3$ (solid lines).
(b) Averaged ratio of acceleration $(a_z - g)/a_x$ with $C_3 =0$ (dashed lines) and $C_3 = 5.6\times10^{-49}$ Jm$^3$ (solid lines).
The black dotted lines represent the threshold velocity to reach the surface.}
\label{azax}
\end{figure}
The calculated results for the $z$-direction integrated-averaged acceleration, along with the average ratio of the $z$-direction to the $x$-direction integrated acceleration, are shown in Fig.~\ref{azax}.
The acceleration along the $z$-direction, normalized by subtracting the gravitational acceleration from the integrated-averaged acceleration in the $z$-direction, is shown in Figure~\ref{azax}(a). 
The dashed line represents the calculation results for evanescent light detuning $\delta$ from +3.4 to +20.2 MHz, assuming that the van der Waals coefficient $C_3 = 0$.
The calculations assumed that the atomic ensemble was initially in thermal equilibrium at a temperature of 22 $\mu$K, centered at 57 $\mu$m. 
The initial velocity distribution aligned with the Maxwell-Boltzmann distribution, with an added initial trap beam velocity of $v_0 = -44$ mm/s prior to release.
The ensemble exhibited a Gaussian width of 65 mm/s.
Velocities ranging from $-$109 to 21 mm/s are expected to contribute significantly to the observed phenomenon.
The black dotted lines in Fig.~\ref{azax} represent the threshold velocity, defined as $v_{\rm th} = -35.1$ mm/s. 
This threshold velocity corresponds to the atoms that reach the surface after a flight time of 2.5 ms. 
Atoms with velocities exceeding this threshold do not reach the surface and therefore do not participate in surface interactions.
Conversely, atoms with velocities below $v_{\rm th}$, in the absence of evanescent light or when evanescent light acts as an attractive potential, are expected to gain kinetic energy upon reaching the surface, dissipate, and subsequently leave the measurement region.
Therefore, only when the evanescent light provides a repulsive potential do atoms with velocities below $v_{\rm th}$ traverse the surface region while interacting, without dissipating from the surface area.
Moreover, when the initial velocity is significantly negative, the repulsive force exerted by the evanescent light alone is insufficient to decelerate the atom, allowing it to reach the surface. 
When $C_3 = 0$, the repulsive potential generated by the evanescent light was the sole force acting on the atom near the surface. Consequently, the effective initial velocity range was broader at $\delta = 6.7$ MHz, where the repulsive force was stronger, whereas the velocity range at $\delta = 20.2$ MHz was narrower than that of the other detunings.
When $C_3 = 0$, atoms in close proximity to the surface can be attracted to and reach the surface owing to the attractive near-field Casimir-Polder potential.
The results for $C_3 = 5.6\times 10^{-49}$ Jm$^3$ are shown in the solid lines in Fig.~\ref{azax}(a).
As the cold atom approached the surface region, it entered the surface region, resulting in a narrower range of initial velocities that can be measured through recapture. 
Furthermore, this measurable velocity range became significantly narrower under conditions of large detuning.

%
%
\begin{figure}
\includegraphics[width=7.0cm]{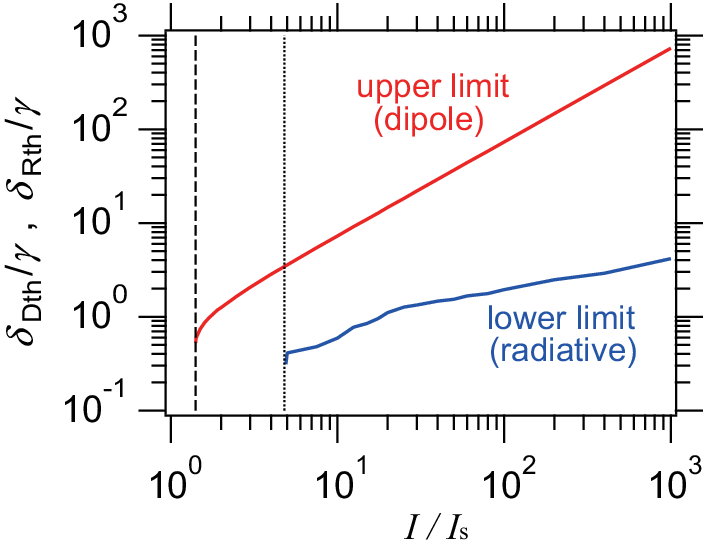}
\caption{Upper limit of detuning $\nu_{\rm Dth}/\gamma$ (red line) for the repulsive force from the evanescent optical dipole force to exceed near-field Casimir-Polder attraction, as well as the lower limit $\nu_{\rm Rth}/\gamma$ (blue line) for the detuning required to maintain the atom below the Rayleigh length of the trap beam during acceleration by the radiative force of the evanescent light.}
\label{delth}
\end{figure}
The ratio of the normalized $z$-direction acceleration in (a) to the $x$-direction acceleration is shown in Fig.~\ref{azax}(b).
In the $x$-direction, acceleration occurred parallel to the surface owing to the radiative force exerted by the evanescent light.
When $C_3 = 0$, the ratio corresponded to $R_{\rm ED}$ within the velocity range exceeding $v_{\rm th}$.
The effect of the optical dipole force relative to the radiative force remains significant, even near resonance, for evanescent light.
In the velocity region below the threshold velocity $v_{\rm th}$, a slight discrepancy from the reference value $R_{\rm DR}$ was observed. 
This deviation could be attributed to the frequency shift induced by the Doppler shift in the surface direction, which becomes more pronounced near the region.
In contrast, when $C_3 = 5.6\times10^{-49}$ Jm$^3$, the acceleration in the $z$-direction was suppressed across the entire velocity range owing to the attractive near-field Casimir-Polder potential.
A larger detuning resulted in an increased lower limit for the initial velocity. 

The upper detuning limit, $\nu_{\rm Dth}/\gamma$, at which the repulsive optical dipole force from the evanescent light overcomes the attractive near-field Casimir-Polder interaction, is shown in Fig.~\ref{delth}. Furthermore, it indicates the lower detuning limit $\nu_{\rm Rth}/\gamma$, which reflects the dependence of the evanescent light radiative force acceleration on the Rayleigh length of the trapping beam (assumed as 1.5 mm in this experiment) for atomic displacement. 
The threshold detuning increased with higher $I/I_s$, and the upper limit set by the optical dipole force increased linearly.
At $I/I_s = 14.6$, the effective range in which the near-field Casimir-Polder force surpassed the optical dipole force at $C_3 = 5.6 \times 10^{-49}$ Jm$^3$ vanished when $\delta_{\rm Dth} > 64.7$ MHz.
In contrast, $\nu_{\rm Rth}/\gamma$ represents the threshold beyond which cold atoms traversed distances exceeding the Rayleigh length. 
Specifically, this threshold corresponds to $\delta_{\rm Rth} = 5.1$ MHz, whereas at $\delta = 3.4$ MHz, a fraction of atoms were propelled beyond the Rayleigh length, where they could not be recaptured.
This accounted for the lower calculated atomic number relative to the experimental value at $\delta = 3.4$ MHz, as shown in Fig.~\ref{expfountaineva}(a), owing to atom loss beyond the Rayleigh length.
Therefore, cold atomic fountain experiments utilizing evanescent light are typically performed near resonance conditions, where the near-field Casimir-Polder potential significantly influenced atomic dynamics.

Atom-surface interactions beyond the mid-range of several hundred nanometers from the surface are dominated by the Casimir-Polder effect, which is proportional to $z^{-4}$ \cite{laliotisAtomsurfacePhysicsReview2021}.
At such distances, the Casimir-Polder potential provides weaker attractive interactions and less deceleration compared with the near-field Casimir-Polder potential, indicating the recapture of more atoms at the surface.
As shown in Fig.~\ref{azax}(a), the velocity range corresponding to the average acceleration narrows as the detuning magnitude increases.
By further reducing the initial velocity spread using atomic ensembles at temperatures lower than those in this study, or Bose-Einstein condensates, atoms that persist for longer durations in regions farther from the surface can be observed selectively.
Under such conditions, the influence of the Casimir-Polder interaction is expected to become more pronounced in the decay curve of the recaptured number of atoms.
Furthermore, in the presence of 
novel
surface potentials---such as the repulsive Casimir-Polder interaction
and the giant Casimir-Polder interaction in Rydberg atoms---the lower limit of the effective velocity range decreased. This potentially enabled the measurement of atoms that have undergone more extensive surface interactions \cite{yuanRepulsiveCasimirPolderPotential2015}.

\section{Conclusions}
The observation of atom-surface interactions was achieved using a cold atomic fountain in conjunction with evanescent light.
For evanescent light detunings ranging from +3.4 to +20.2 MHz, the number of recaptured atoms increased, and the calculated results incorporating the near-field Casimir-Polder (van der Waals) potential $C_3$ demonstrated strong agreement with the experimental data.
A comparative analysis between available $C_3$ calculations and the experimental results was conducted, with the least-squares error being considered. 
This analysis yielded $C_3 = 5.6^{+2.4}_{-1.9}\times10^{-49}$ Jm$^3$. 
$In situ$ calculations for the cold atomic fountain demonstrated that the range of initial velocities permitting Casimir-Polder interactions was influenced by both $C_3$ and the detuning $\delta$. 
Overall, this study elucidated the experimental parameters necessary for observing surface interactions using near-resonant evanescent light and a cold atomic fountain.

\begin{acknowledgments}
We gratefully acknowledge K. Shibata, S. Asahi, Y. Kobayashi, and G. Tanaka for their valuable assistance during the construction of the initial experimental setup.
This work was supported by the Matsuo Foundation, Research Foundation for Opto-Science and Technology, Chuo University Joint Research Grant, Chuo University Grant for Special Research, and JSPS KAKENHI Grant Numbers JP23K03287 and JP23K25893. 
\end{acknowledgments}

\newpage


\bibliography{biblio_coldeva}



\end{document}